# Securing Virtual Reality Experiences: Unveiling and Tackling Cybersickness Attacks with Explainable AI

Ripan Kumar Kundu, Matthew Denton, Genova Mongalo, Prasad Calyam
and Khaza Anuarul Hoque, *Senior Member, IEEE*

*Abstract*—The synergy between virtual reality (VR) and artificial intelligence (AI), specifically deep learning (DL)-based cybersickness detection models, has ushered in unprecedented advancements in immersive experiences by automatically detecting cybersickness severity and adaptively various mitigation techniques, offering a smooth and comfortable VR experience. While this DL-enabled cybersickness detection method provides promising solutions for enhancing user experiences, it also introduces new risks since these models are vulnerable to adversarial attacks; a small perturbation of the input data that is visually undetectable to human observers can fool the cybersickness detection model and trigger unexpected mitigation, thus disrupting user immersive experiences (UIX) and even posing safety risks. In this paper, we present a new type of VR attack, i.e., a cybersickness attack, which successfully stops the triggering of cybersickness mitigation by fooling DL-based cybersickness detection models and dramatically hinders the UIX. Next, we propose a novel explainable artificial intelligence (XAI)-guided cybersickness attack detection framework to detect such attacks in VR to ensure UIX and a comfortable VR experience. We evaluate the proposed attack and the detection framework using two state-of-the-art open-source VR cybersickness datasets: Simulation 2021 and Gameplay dataset. Finally, to verify the effectiveness of our proposed method, we implement the attack and the XAI-based detection using a testbed with a custom-built VR roller coaster simulation with an HTC Vive Pro Eye headset and perform a user study. Our study shows that such an attack can dramatically hinder the UIX. However, our proposed XAI-guided cybersickness attack detection can successfully detect cybersickness attacks and trigger the proper mitigation, effectively reducing VR cybersickness.

## I. Introduction

Virtual Reality (VR) has gained immense popularity in many application domains, including education [1], military [2], health care [3], gaming [4], social networking [5] and many more. Unfortunately, with their increasing popularity, safety, security, and privacy risks of VR have also been reported and demonstrated by several researchers [6]–[9]. For instance, the authors in [10] perform a Keylogging attack from VR head motion, whereas the Keylogging was performed through a wifi side-channel attack in [11]. The impact of VR malware and Man-in-the-Room (MitR) attacks was demonstrated in [5]. An interesting attack where the VR user is forced to hit a wall or object, known as the Human Joystick attack, was demonstrated in [12]. User experience is a vital factor in VR. Interestingly, there are only a few works that utilize this opportunity to attack VR users. For instance, the authors in [1], [13] use network, GPU vulnerabilities, and malicious scripts to trigger user experience attacks through cybersickness in VR users. Cybersickness is a set of unpleasant symptoms such as eye strain, dizziness, headache, disorientation, and such [14]–[16] that pose a severe threat to the immersive experience of the user.

Applying cybersickness mitigation without assessing the users' state using the cybersickness detection method may hinder users' immersive experience. Moreover, many existing cybersickness detection methods are not automated and require human intervention for detecting and measuring them, which also makes triggering the appropriate cybersickness migration unrealistic [17]–[20]. Motivated by this, machine/deep learning (ML/DL)-based automated methods for detecting cybersickness have gained much attention from the researchers [21]–[29]. In an ideal setup (with no adversary), the outcome of such state-of-the-art (SOTA) ML/DL-based cybersickness detection can be used to trigger different cybersickness mitigation methods automatically (e.g., manipulating the field of view (FOV), changing the depth of field (DOF) blur, etc.) [17], [30], [30]–[34]. However, ML/DL algorithms can also be fooled by carefully crafted adversarial examples [35]. This means that if an attacker can successfully fool the cybersickness DL model, the cybersickness mitigation will not be triggered. We utilize this idea to develop a new type of VR attack, i.e., *cybersickness attack* by manipulating the inherent vulnerability of cybersickness detection models. *Let us consider a scenario where a VR user, Bob, a gaming enthusiast, is playing a VR game with a built-in DL-enabled cybersickness prediction and mitigation system to ensure a smooth and comfortable VR adventure. However, during his latest VR adventure, he unexpectedly suffers from severe cybersickness. Despite the cybersickness prediction and mitigation ability of the VR headset, the user had an uncomfortable experience, which led him to exit the VR simulation. After repeated complaints, the VR headset company started an investigation that revealed that an anonymous attacker exploited the cybersickness prediction model's vulnerability, manipulating user data to fool the outcome of the prediction algorithm.* Such attacks target the prediction algorithm's ability to detect and prevent cybersickness, leading to distorted experiences for users like Bob, and thus can erode trust in VR systems, potentially hindering widespread adoption

Manuscript received July 19, 2024; revised February 16, 2025. This material is based upon work supported by the National Science Foundation (NSF) under Award Numbers: CNS-2114035. Any opinions, findings, conclusions, or recommendations expressed in this publication are those of the authors and do not necessarily reflect the views of the NSF.

and breaking user immersion, diminishing engagement with the virtual environment.

To bridge the research gap in adversarial attack detection for SOTA ML/DL-based cybersickness detection models, in this paper, we propose a novel framework explainable artificial intelligence (XAI)-guided adversarial attack detection in DL-enabled cybersickness model in VR for ensuring user immersive experience. Our proposed framework differentiates between normal and adversarial examples of cybersickness prediction through the use of SHapley Additive explanations (SHAP)-based post-hoc explanations techniques. Our contributions can be summarized as follows:

- First, we develop a baseline system with cybersickness detection and mitigation capabilities. For accurate cybersickness detection, we develop and evaluate the performance of three different DL models: convolutional neural network-long short-term memory model (CNN-LSTM), LSTM, and gated recurrent unit (GRU). To train these models, we use two SOTA open-source VR cybersickness datasets: Simulation 2021 [26], [36] and Gameplay [37], [38] dataset. Once the cybersickness is detected, we mitigate using *GingerVR* [33], an open-source software package for the Unity game engine. Specifically, we use three mitigation techniques from GingerVR: dynamic Gaussian blurring, dynamic field-of-view (FOV), and foveated depth-of-field (DOF) blur based on the user cybersickness severity levels. We show that this system works well, as expected, with cybersickness detection and mitigation in an ideal environment, i.e., when no cybersickness attack is launched.
- Next, to launch a cybersickness attack in VR, we generate adversarial examples and inject them into the cybersickness detection models to fool the detection outcome. To craft the adversarial examples, we utilize and evaluate the impact of three well-known adversarial example generation algorithms: Fast Gradient Sign Method (FGSM), Projected Gradient Descent (PGD), and Carlini-Wagner (C&W). We evaluate our attacks in both black-box and white-box settings. We show that our proposed cybersickness attack is capable of successfully fooling the system. For instance, the C&W attack causes a $5.69\times$ and $4.65\times$ decrease in accuracy for the LSTM-based cybersickness detection model on the Simulation 2021 and Gameplay datasets, respectively, compared to the accuracies without the cybersickness attacks. This successfully stops the triggering of cybersickness mitigation, and our user study shows that this dramatically hinders the UIX.
- We develop an ML and DL-based binary adversarial example detection method to detect cybersickness attacks. This is achieved by differentiating XAI-generated *attack signatures* from the cybersickness detection models. Specifically, to generate the attack signatures, we use the SHapley Additive explanations (SHAP)-based post-hoc explanations techniques. We develop and evaluate one DL model, namely feed-forward neural network (FFNN), and two ML models, namely random forest (RF), and XGBoost (XGB), for the detection of cybersickness attacks. We show that our proposed method can successfully detect cybersickness attacks with an accuracy of 90% ( Simulation 2021 dataset) and 94% (Gameplay dataset).
- Finally, to verify the effectiveness of our proposed method, we built a VR testbed to conduct a user study on cybersickness attack detection and proper mitigation trigger evaluation, in which the users were immersed in a custom-built VR roller coaster simulation with an HTC Vive Pro Eye headset. Our user study results show that the proposed method successfully detects cybersickness attacks and triggers the proper mitigation method, effectively reducing cybersickness in VR users, improving UIX, and ensuring a comfortable VR experience. It is important to note that the proper mitigation is triggered by discarding the adversarial sample and only forwarding the normal sample to the mitigation engine.

To the best of our knowledge, this work is the first work that attacks the UIX by manipulating the cybersickness detection model's vulnerability and providing an effective defense against that.

## II. Related Works

Recently, VR has become a target of attackers from various perspectives. Prior studies have identified and explored different cybersecurity risks in VR, such as the human joystick attack, overlay attack, side-channel attack, security & privacy attack, denial-of-service attack (DoS), frame-rate-oriented attack, virtual-physical perceptual manipulations attack, etc., [1], [9], [11]–[13], [39], [40] which disrupt the user experience during VR immersion and induce cybersickness either directly or indirectly. For instance, authors in [1] applied network-based security privacy attacks (SP) in cloud-based Virtual Reality Learning Environments (VRLEs). They found that these SP attacks lead to disruption of the user immersive experience (UIX) (e.g., disruption of learning and disruption of user safety such as physical harm and cybersickness) and adversely impact the application's usability. In contrast, Blessing et.al. [13] applied GPU-based and network-based attacks to cause visual, interaction, locomotion, and auditory disruptions and cybersickness in VR users. Indeed, cybersickness causes severe discomfort such as eye strain, nausea, disorientation, and many more [41] while using head-mounted displays. Therefore, a plethora of research has been conducted for cybersickness detection and mitigation. To measure cybersickness in real-time, researchers have proposed different ML and DL-based methods to detect cybersickness automatically from a variety of data such as Simulator Sickness Questionnaire (SSQ) data, different objective measurements (e.g., heart rate (HR), eye-blink rate, pupil diameter, etc.,) in HMDs [21]–[24], [26]–[29]. For instance, Islam et al. [24], [26] proposed a multimodal deep fusion network for classifying cybersickness from the multimodal integrated sensors data. In addition to cybersickness detection, an adaptive VR-based approach has been proposed to automatically detect cybersickness severity during immersion and dynamically improve the experience by applying several visual techniques, such as dynamic field of view (DFOV), depth of field (DOF) blur, etc., that help reduce cybersickness during immersive experiences without any interventions [17], [30], [34], [42]. Unfortunately, their work did not consider the security vulnerabilities related to these DL methods, which pose a significant threat to the upcoming VR ecosystem (e.g., Metaverse). In VR, cybersickness mitigation depends on accurate cybersickness detection. Moreover, DL algorithms can also be fooled by carefully crafted adversarial examples [35].

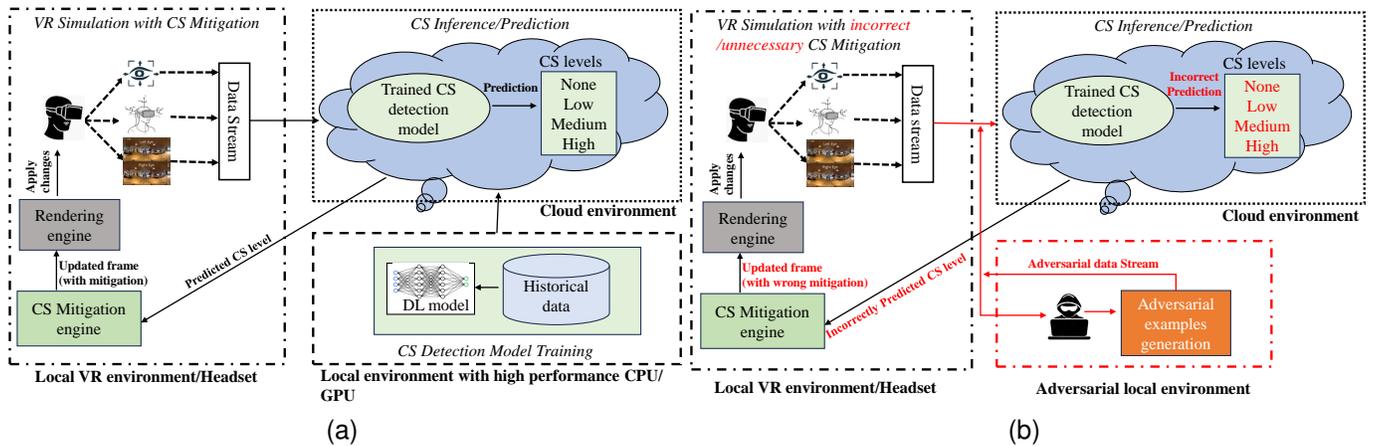

Figure 1: (**a**) Cybersickness detection and mitigation in an ideal scenario (without cybersickness attack), and (**b**) Cybersickness detection and mitigation under adversarial attack conditions (with cybersickness attack).

Adversarial attacks mainly rely upon the fact that any relatively complex ML/DL model is a black box making it impossible for human observers to explain the process the model uses to make its decisions. However, through XAI, users can gain insights into how a DL model makes decisions regarding cybersickness. Prior works have applied XAI-based post-hoc explanation methods for explaining the cybersickness prediction model to identify the most dominating features for causing cybersickness; however, their works are limited to feature explanation and model size reduction [27], [28]. We hypothesize that a deep connection exists between model explainability and adversarial examples. Intuitively, a well-explained model should be robust to adversarial perturbations since adversarial input would result in anomalous explanations for the model's decision. Furthermore, it should be noted that adversarial examples are an intrinsic property of the data itself, which is categorized as robust and non-robust features [43]. Non-robust features are highly predictive yet fragile and prone to change drastically due to small input perturbations. In contrast, robust features are highly predictive features and do not change easily by small changes in the input. Thus, we hypothesize that we should see different patterns in the importance of robust vs. non-robust features in the cybersickness classification of normal and adversarial inputs by utilizing XAI methods for interpreting model predictions. Unfortunately, no studies have attempted to use XAI for adversarial example detection in DL-enabled cybersickness prediction in VR, which motivates our work.

## III. BASELINE CYBERSICKNESS DETECTION AND MITIGATION

This section presents the details of our baseline cybersickness detection and mitigation in an ideal scenario (without cybersickness attack). As shown in Figure 1 (a), It is comprised of two phases: (1) Baseline cybersickness detection using deep learning and (2) Cybersickness mitigation phase development. The details of these processes are described below.

### A. Cybersickness Detection using Deep Learning

To develop the baseline system for accurate cybersickness detection, we use three DL models: long short-term memory (LSTM), grated recurrent unit (GRU), and convolutional neural network (CNN)-LSTM. The LSTM model has six LSTM layers with size 128, one Dropout layer with a recurrent dropout of 15% to reduce overfitting, and two dense layers with the rectified linear unit (ReLU) activation function. The CNN-LSTM model is composed of two convolution layers followed by two fully connected layers, each with a filter size of 64, kernel size of 3, one max pooling layer with pooling size of 2, and one LSTM layer with size 128, which is then passed through the ReLU, one dropout layer with a dropout rate of 15%, and two dense layers with the activation function. The GRU model has three GRU layers with sizes 32, 64, and 128, where each layer is followed by a dropout layer with a dropout of 20% and two dense layers with the ReLu activation function. We chose these DL models as they can capture datasets' temporal and spatial features. For both of these DL models, we set a timestep of 90. We use 'softmax' as the activation function in the last dense layer, which contains multiple outputs for the multiple classes of cybersickness. We use categorical cross-entropy as the loss function in our DL-based cybersickness prediction models.

### B. Datasets

We train the above-mentioned baseline models using the Simulation 2021 dataset [26], [36] and Gameplay dataset [37], [38] for the cybersickness classification. The details of these two datasets are given in the following subsections. It is worth mentioning that there is a lack of open-source cybersickness detection datasets. Thus, SOTA works in automated cybersickness detection often relies on Simulations 2021 (also known as integrated sensor dataset) and Gameplay datasets [23], [28], [29], [37], [44].

*1) Simulation 2021:* The Simulation 2021 dataset [24], [36] contains eye tracking, head tracking, and physiological data collected from 27 participants (Male: 15 and Female: 12) with a mean age of 30.04 years and diverse ethnic backgrounds, such as Asian, Black, Caucasian, Hispanic, and Pacific Islander. The participants were immersed in five VR simulations:

Roller Coaster, Roadside, Beach City, Sea Voyage, and Furniture Shop. The eye tracking, head tracking, and physiological data are each subdivided into multiple categories. The subcategories in the eye tracking data are Convergence Distance, Gaze Direction (x, y, z), Pupil Diameter (left), Pupil Position (x, y, z), and % of Eye Openness. The head tracking data consists of Quaternion Rotation for the X, Y, Z, and W axis. The dataset contains four cybersickness severity categories: none, low, medium, and high. Note, we are interested in predicting cybersickness and applying adversarial attacks from the integrated sensor data readily available in many current HMDs; we exclude the physiological data ( e.g., heart rate (HR) and electrodermal activity) from the dataset. The reason is that HR and electrodermal activity data are typically recorded through the external sensors, which limits VR locomotion and 3D-object manipulation during the immersion and are not conducive to room-scale VR experiences (e.g., Beat Saber).

*2) Gameplay:* The Gameplay dataset [4] contains 22 different features, such as candidate profiles, questionnaires, user field of view, speed of the game in playtime, etc., for 35 participants. This dataset is generated using two VR games: a racing and a flight game. The study originally contained 87 subjects. However, only the data from 35 participants (Male: 26, Female: 9) was collected since the study only collected data from participants who answered all virtual reality subjective questionnaires (VRSQ) correctly and completed the whole VR session. The dataset contains four cybersickness severity classes: *none*, *slight*, *moderate*, and *severe*. It is worth mentioning that for uniformity, we rename these four classes as follows– *none*: none, *slight*: low, *moderate*: medium, and *severe*: high to match the Simulation 2021 dataset.

*C. Cybersickness Mitigation Phase Development*

Once the severity of cybersickness is flagged (e.g., low, medium, and high), it is forwarded to the mitigation engine via an HTTP request. Then, the mitigation method is applied based on that severity. However, no mitigation will be triggered if the detected cybersickness level is *none*. Finally, the mitigation method is forwarded to the rendering engine to update the frame. In this work, we use GingerVR [33] to develop the mitigation engine. GingerVR is an open-source software package for the Unity game engine that contains several cybersickness mitigation techniques. However, we restrict ourselves to three popular cybersickness mitigation techniques for developing the mitigation library: dynamic Gaussian blurring, dynamic field-of-view (FOV), and foveated depth-of-field (DOF) blur. Dynamic Gaussian blurring uses blur to reduce the amount of optical flow available to the user [45]. On the other hand, the dynamic FOV mitigation technique puts a virtual filter to block the corners of their vision and change its center based on the user's eye gaze position [19]. However, the foveated DOF blur uses ideas from foveated imaging and DOF to create a scene with no artifacts, which helps in reducing cybersickness [17]. The goal of this paper is not to propose a new method of cybersickness mitigation but instead to utilize the already established method. Our choices of cybersickness mitigation are inspired by the literature such as [34], which demonstrates the varying effectiveness of these techniques at different severity levels.

The mitigation engine automatically triggers the cybersickness reduction technique to match the intensity of the severity values (e.g., low, medium, and high) detected by the cybersickness detection model. For instance, the mitigation engine triggers the foveated depth-of-field (DOF) blur technique once the cybersickness level is predicted as low. This technique has previously shown that users could reduce mild cybersickness conditions through the visual changes introduced by the foveated DOF blur technique [17], [46]. On the other hand, the mitigation engine triggers the dynamic Gaussian blur technique if the predicted cybersickness level is medium. By applying the dynamic Gaussian blur technique at this stage, the mitigation engine reduces cybersickness by minimizing visual stimuli and motion perception [30], [34]. However, applying such a technique at higher levels of cybersickness may introduce further visual disruptions, potentially increasing user discomfort instead of providing significant relief [33], [34]. This is because when users are already distressed at higher levels of cybersickness, adding further visual alterations could intensify discomfort and exacerbate their condition [17], [47], [48]. In such cases, i.e., if the predicted cybersickness level is high, the mitigation engine triggers the dynamic FOV reduction technique, which can be adjusted based on the user movement speed [34], [49]. Users at low or mid levels of cybersickness may still be able to tolerate the entire field of view, and prematurely applying FOV reduction could lead to a less immersive experience [34], [42], [50], [51]. Thus, limiting the user's field of view, i.e., applying the DFOV technique, reduces the visual stimuli and motion cues contributing to cybersickness for users with high levels of cybersickness. It is important to note that the rationale behind selecting different mitigation techniques for varying levels of cybersickness stems from each method's distinct advantages and limitations when applied to different levels of cybersickness severity. Rather than relying on a single technique (e.g., dynamic FOV), our approach combines multiple methods to balance cybersickness mitigation and user experience. While dynamic FOV effectively reduces severe symptoms, it can disrupt immersion by limiting the visual field and creating a tunnel vision effect [20], [49]. Less intrusive techniques, such as dynamic Gaussian blur and foveated DOF blur, are better suited for low and medium levels of cybersickness severity, providing relief without significantly narrowing the field of view. A multi-technique cybersickness mitigation approach, tailored to different severity levels, ensures a balanced solution that optimizes both cybersickness mitigation and user experience, as supported by prior studies and practical considerations.

IV. PROPOSED CYBERSICKNESS ATTACKS IN VR

In this section, we describe the details of the proposed cybersickness attack, including their threat models, attack scenarios, and methods for crafting them.

*1) Formalization of the problem:* The sensors in VR headsets, such as eye-tracking and head-tracking, record measurements typically at periodic time steps, constituting multivariate time series (MTS) data. This VR MTS data from $N$ sensors can be denoted as $X = [x_1, x_2, ..., x_T]$, where $T = |X|$ represents the length of $X$ and $x_i \in \mathbb{R}^N$ is a $N$ dimensional data point at time $i \in [1, T]$. Let us further assume that $D = (x_1, Y_1), (x_2, Y_2), ..., (x_T, Y_T))$ is the dataset of pairs $(x_i, Y_i)$, where $Y_i$ is the set of corresponding one-hot label vectors (representing cybersickness class (e.g., 'none', 'low', 'medium', and 'high')) corresponding to $x_i$. Given this, the MTS cybersickness classification task consists of training a model with dataset D to map from the space of possible inputs to a probability distribution over the class variable values (labels). Now, Let $f(\cdot) : X \to Y$ represents a DL model for cybersickness classification. Furthermore, let $X'$ denotes the adversarial example, a perturbed version of $X$ (the original instance) such that $Y \neq Y'$ and $\|X - X'\| \leq \varepsilon$, where $\varepsilon$ denotes the maximum perturbation magnitude.

Thus, given a trained DL classifier $f$ and an input $X$, crafting an adversarial example $X'$ can be described as the following box-constrained optimization problem: $\min_{X'} \|X' - X\|$ s.t. and $f(X') = Y'$, $f(X) = Y$ and $Y \neq Y'$

*2) Threat model for the VR cybersickness attacks:* This work considers both a *white-box* and a *black-box* attack scenario. In both cases, we do not perturb the cybersickness detection model parameters. We use only the crafted adversarial examples to fool the cybersickness detection. In the white-box attack scenario, we assume that the adversary has the full knowledge of the cybersickness detection models, which helps craft powerful adversarial examples. In contrast, in the black-box attack scenario, the adversary is assumed not to have access to the model's internal parameters. Instead, the attacker can make an informed guess about the cybersickness detection models for crafting adversarial examples. We demonstrate the applicability of black-box attacks by exploiting the transferability [52] property of the adversarial examples.

**Attack objective:** The attacker aims to fool the VR cybersickness detection method by misclassifying the user's detected cybersickness level. This may lead to two situations: (1) VR cybersickness mitigation does not trigger when required, e.g., the cybersickness detection model classifies the user's cybersickness level as none preventing the triggering of cybersickness mitigation, whereas the user is feeling severe cybersickness, and (2) Wrong type of VR cybersickness mitigation is triggered that does not match with the cybersickness severity level, e.g., the user is feeling slight cybersickness; however, the prediction model is predicting severe cybersickness, or the user is not feeling cybersickness, the model predicts the user's cybersickness level as a severe triggering unwanted intense cybersickness mitigation method. Table I shows how the unnecessary/incorrect mitigation technique will be triggered under the cybersickness attack condition. Nevertheless, in all these cases, the outcomes will severely hamper users' UIX and may force them to quit the VR simulation.

**Attack surface:** To gain access to the VR headset, the cloud where the cybersickness detection model is stored, or intercepts the communication between the headset and the cloud server, allowing malicious code to be executed. The adversary could achieve this through various attack methods such as spoofing, phishing, evil twin attacks, or other social engineering attacks. For example, through phishing or social engineering, the adversary could trick the user of the cloud environment into downloading a malicious program that, when executed, copies the cybersickness detection model to the adversary's device. Another possible method by which the adversary may gain code execution privileges is exploiting the security flaws of the software, firmware, or hardware of either the user's device or the cloud environment's infrastructure [53]–[55]. After that, the attacker observes the measurements of the VR sensors, such as eye tracking, head-tracking, etc., and feeds the crafted adversarial examples. For instance, Casey et al. [12] showed that an attacker forced VR users into hitting physical objects and walls by activating their HMD cameras without their awareness. Furthermore, authors in [56] presented that passwords entered in a virtual environment while using the Samsung Gear VR headset can be inferred by exploiting the side channel information of users' activities. They used computer vision-based and motion sensor-based models to extract the password and manipulate the user's HMD tracking system to generate a fake sync pulse [57]. In contrast, authors in [58] demonstrated that by using an application running concurrently on a user's GPU, an attacker could monitor the GPU's usage to recover the parameters of an ML model with considerable accuracy. Using such techniques, the attacker can inject the crafted adversarial examples, resulting in a disconnect between the user's interactions with the environment and their visual stimuli, leading to unexpected cybersickness. Note that even if an attack detection method is deployed in the cloud or VR headset, they may be unable to distinguish between false and legitimate VR sensor readings since the (adversarial) attack signature can be very stealthy. An overview of the considered attack scenario for the cybersickness detection and mitigation scheme is shown in Figure 1.

Table I: Cybersickness attack objective with consequences.

| Benign CS level (no attack) | Target CS level (under CS attack) | Attack consequence |
|---|---|---|
| None (N) | (L, M, H) | Unnecessary mitigation |
| Low (L) | (N, M, H) | Incorrect mitigation |
| Medium (M) | (N, L, H) | Incorrect mitigation |
| High (H) | (N, L, M) | Incorrect mitigation |

*3) Adversarial example generation for VR cybersickness attacks:* We utilize three popular and effective algorithms: Fast gradient sign method (FGSM) [59], Projected gradient descent (PGD) [60], and Carlini-Wagner (C&W) [61] to generate adversarial examples for VR cybersickness attacks. It is worth mentioning that adversarial attacks from the computer vision domain cannot be directly applied to the VR sensors due to the nature of the attacks (pixels vs. time-series classification). Moreover, in computer vision, small perturbations are added to the images' pixels, leading to the misclassification of the images. In contrast, the attackers' objective in the VR domain is to add small perturbations to VR sensor-obtained measurements, leading to erroneous cybersickness prediction. Thus, we adopted the adversarial attack algorithms from the computer vision domain for time-series data and implemented them to suit the context of VR sensor inputs, inspired by [62], [63]. Unlike traditional adversarial attacks focused on static image data, our method targets time-series data from VR eye-tracking and head-tracking data inherent in VR systems, requiring consideration of temporal dependencies and the sequential nature of the data. Additionally, the attacks are designed to operate during immersion to trigger unexpected mitigation techniques.

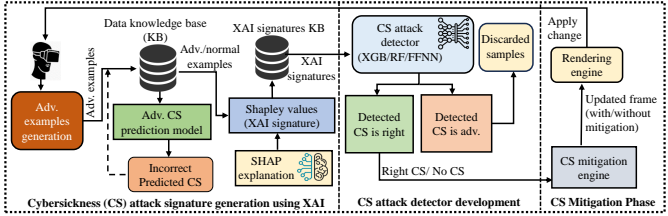

Figure 2: Overview of proposed cybersickness attack defense and mitigation framework.

*4) VR cybersickness attack execution:* Once the adversary gains access to the VR device, cloud where the cybersickness detection model is stored, or successfull in intercepting the communication between the headset and cloud server, the attacker monitors VR sensor time series data, such as eye-tracking and head-tracking measurements, and uses them to craft adversarial examples into the data stream carefully by adding small, strategically calculated perturbations to the time series data using adversarial attack algorithms as presented in Section IV-3 and inspired by [62], [63]. These perturbations are calculated to be imperceptible to users but sufficient to mislead the DL-enabled cybersickness detection model. The crafted adversarial examples are then fed into the cybersickness detection model (instead of the original sensor measurements) during runtime by intercepting the VR eye-tracking and head-tracking sensor time series data stream. If successful, these adversarial inputs may cause the model to output incorrect predictions of cybersickness severity level, preventing appropriate mitigation or triggering unnecessary mitigation techniques. For instance, slight alterations in the temporal patterns of head-tracking data can shift the model's output. When the model predicts a low cybersickness level based on smooth head movements, introducing minor, rapid fluctuations into the head-tracking data can deceive the model into predicting a higher sickness level, prompting unnecessary mitigation measures.

## V. DEFENDING CYBERSICKNESS ATTACK

This section describes a novel defense method against the proposed cybersickness attack mechanism. As illustrated in Figure 2, it is comprised of three phases: (1) Cybersickness attack signature generation using XAI, (2) Cybersickness attack detector development, and (3) Cybersickness mitigation phase. The details of these processes are described below.

*1) Cybersickness attack signature generation using XAI:* An identified cybersickness event in VR can be a legitimate prediction by the cybersickness detection model or an adversarial deception using the proposed attack. For effective cybersickness attack detection, we need to train the detector to differentiate between benign vs. adversarial cybersickness decisions through feature relevance explanations associated with the input. Such explanations indicate each feature's influence on the model's final prediction. The underlying assumption is that the model's response to benign and adversarial samples differs, even if they look similar to humans. As a consequence, the model should come up with very different explanations for these two types of samples [64]. Thus, we use the SHAP-based post-hoc explanation technique to generate the cybersickness attack signatures. SHAP is a feature importance explanation approach that assigns a feature significance value to each prediction. It is based on the mathematical foundation of Shapley values from cooperative game theory [65]. For a given set of input samples (e.g., eye-tracking and head-tracking data) and DL-based cybersickness models (e.g., CNN-LSTM, GRU, and LSTM), the goal of SHAP is to explain the predicted cybersickness of input samples by calculating the contribution of each feature to the prediction. In other words, SHAP values explain how a given feature value increases or decreases a model's prediction. For instance, given a cybersickness prediction model that predicts whether a sample has cybersickness, the SHAP explanation allows us to know to what extent a feature drove a given prediction. Such explanations can be global or local. The overall feature importance ranking (global explanation) of cybersickness can be visualized using SHAP in terms of bar graphs. On the contrary, for the local explanation, each sample is randomly chosen from the test dataset, which contains all the features and determines which features increase or reduce the likelihood of cybersickness for a specific example. For our case, we use the *Deep SHAP* method [66] for computing SHAP values from the DL-based cybersickness models. More specifically, we apply SHAP to the penultimate layer of the cybersickness prediction model to compute these values, similar to the work [64]. After that, these SHAP values are used to calculate the feature's importance, denoted as *XAI signatures*, during cybersickness severity prediction. Then, we use these XAI signatures as features for our adversarial example detector. Next, we compile these XAI signatures to create a dataset denoted as a *repository* in which each signature is labeled either "0" for a benign or "1" for an adversarial example. We repeat this for each model on each dataset. It is worth mentioning that this repository is updated with the XAI signature whenever an attack is detected for enhanced detection performance. Also, note that to generate the attack signature and update the repository, we use a local explanation method, in which each sample is selected from the dataset, which contains all the features and determines which features increase or reduce the likelihood of cybersickness. In this work, the explanation generation process is offloaded to the cloud, where the SHAP-based explanations are computed to generate the XAI signature. This offloading ensures that the real-time VR experience remains smooth and free from the computational cost of SHAP while benefiting from its interpretability and robustness. Additionally, this approach allows for scalable processing power, optimizing the trade-off between defense accuracy and real-time performance.

*2) Cybersickness attack detector development:* Using the XAI signatures described in the previous section, we train two ML models (XGBoost and Random Forest) and one DL model (Feed Forward Neural Network) to detect cybersickness attacks. The overall method is illustrated in Algorithm 1.

The algorithm takes training and testing XAI signatures datasets as input and returns the flagged cybersickness attacks. As shown in Algorithm 1, we first train `TrainAttackDetector()` function to discriminate between normal and adversarial samples. To do this, at first, the

**Algorithm 1:** XAI-guided cybersickness attack detection

| | |
|---|---|
| **Inputs** | : XAI signature training dataset = $S_{tr}$, XAI signature testing dataset = $S_{ts}$ |
| **Outputs** | : Verdict = $\{True, False\}$ |

1: $D_A$ = TrainAttackDetector($S_{tr}$)
2: **for each** sample $i \in S_{ts}$ **do**
3:    Verdict = AttackDetection($D_A$, $S_{ts}$)
4:    **if** Verdict == True **then**
5:       Display message on VR "cybersickness attack detected";
6:       **Continue**;
7:    **end if**
8: **end for**

XAI signature training dataset $S_{tr}$ is fitted with the training function `TrainAttackDetector()` (Line 1). Next, the attack is detected from the fitted trained model $D_A$ based on the attack detection function `AttackDetection()` using XAI signature testing dataset $S_{ts}$ (Line 3). The `AttackDetection()` function calculates the probability of $D_A$ between 0 (*False*) and 1 (*True*) of the XAI signatures, in which normal samples are denoted as right cybersickness $Verdict = False$ (no cybersickness attack) and adversarial samples are denoted as wrong cybersickness $Verdict = True$ (cybersickness attack). Finally, the algorithm detects normal and adversarial samples when the attack detection process is complete (Lines 4–9). Once the attack is detected, it sends an alert to the user. It is important to note that once the attack is detected, the adversarial sample is discarded, and only the samples without attacks are forwarded to the mitigation engine to trigger the proper mitigation technique.

*3) Cybersickness mitigation phase:* Once no cybersickness attack is detected, denoted as normal samples, it is forwarded to the mitigation engine via an HTTP request. Then, the mitigation method is applied based on the severity of the cybersickness if these two conditions are met: (i) the cybersickness detector flags the sample as normal (without attack), and (ii) the detected cybersickness level is either low, medium, or high, as discussed in Section III-C.

## VI. EXPERIMENTAL SETUP

This section explains the experimental setup, performance metrics, and hyper-parameter we used to validate our proposed XAI-guided cybersickness attack detection framework. We used Scikit-Learn [67] and TensorFlow-2.4 [68] for training and evaluating our DL models. We used the CleverHans library [69] to implement cybersickness attacks. For explaining the DL models and generating "XAI signatures", we used the SHAP [65] library. We used Scikit-Learn [67] for training the ML and DL-based cybersickness attack detectors. All of the models were trained on an Intel Core i9 Processor and 32GB RAM option with NVIDIA GeForce RTX 3080 Ti GPU.

**Performance Metrics:** The performance of the DL-based cybersickness classification can be quantified using standard quality metrics such as accuracy, precision, recall, and F-1 score [27]. The adversarial attacks are evaluated using accuracy, precision, recall, and F-1 score metrics similar to the work [64]. Furthermore, the cybersickness attack detection performance is evaluated using accuracy and F1-score metrics.

**Hyper-Parameter:** We used the Adam optimizer to optimize our DL models with epochs of 200 and a batch size of 256. For training the DL models, we used the learning rate of 0.001. To prevent the model from overfitting, we deployed an early-stopping strategy with a patience value of 30 while training the DL models. Furthermore, to train the ML-based cybersickness attack detectors, we used the selected hyperparameters from Scikit-learn [67] in which every model is trained with individual hyperparameter values. For instance, the number of estimators for RF and XGB classifiers is 30, and 40 and the learning rate for the XGB classifiers is 0.05, respectively. For training the DL-based cybersickness attack detector (FFNN model), we use the learning rate of 0.001 with epochs of 100 and a batch size of 64.

## VII. SIMULATION-BASED EXPERIMENTAL RESULTS

This section presents the simulation-based experimental results of the proposed framework for cybersickness detection, cybersickness attack impact, and XAI-guided defense in VR.

### A. Baseline Cybersickness Detection and Explanation

This section presents the results of baseline models of cybersickness detection and explanation.

*1) Baseline cybersickness detection using LSTM, GRU, and CNN-LSTM models:* Table II summarizes the accuracy, precision, recall, and F-1 scores of cybersickness severity classification using LSTM, GRU, and CNN-LSTM models for Simulation 2021 and Gameplay datasets under normal conditions. The cybersickness classification using the CNN-LSTM, LSTM, and GRU model resulted in an accuracy of 90%, 91%, and 90% for the Simulation 2021 dataset and 94%, 93%, and 92% for the Gameplay dataset, respectively. It is observed that the overall performance of the LSTM model is slightly better than the GRU and CNN-LSTM model regarding precision, recall, and F1-score for cybersickness classification.

*2) Baseline cybersickness global and local explanations:* We apply the SHAP-based global and local explanation method to explain the DL-based cybersickness model's prediction outcome. For the global explanation, the overall feature importance for cybersickness severity classification using the LSTM model and the CNN-LSTM model for the Simulation 2021 and Gameplay dataset is visualized in Figure 3. From Figure 3 (a) and (b), we observe that for the Simulation 2021 dataset, features such as *NrmLeftEyeOriginY* and *NrmLeftEyeOriginZ*, meaning the normalized left eye origin on the Y and Z axis, *Right Eye Openness*, meaning the right eye is open as a percentage, *GazeOriginWrldSpcY*, meaning that gaze origin in world space Y axis etc., are the most influential features for cybersickness severity classification under normal condition. It is observed that eye-tracking features have a much more significant impact on the cybersickness severity classification than head-tracking features. The reason is that eye tracking features contain insightful information such as the type of blink of the user, gaze behavior, and the position of the pupil to track the user's activity. Similarly, from Figure 3(c) and (d), it is observed that features such as *Camera Auto Movement*, *Camera Field Of View*, *StaticFrame*, *User Eye Dominance*, etc., have a much stronger influence in the cybersickness classification for the Gameplay dataset.

The results of the local explanation utilizing SHAP for the LSTM model for the Simulation 2021 and Gameplay datasets

Table II: Mean accuracy, precision, recall, and f1-score of cybersickness prediction without attack models, and with FGSM ($\varepsilon = 0.1$), PGD ($\varepsilon = 0.1$, $\alpha = 0.01$, and $I = 20$), and C&W (max iterations = 1000) attacks cybersickness models. The notation X / Y represents the percentage of the Simulation 2021 dataset/Gameplay dataset.

| Attack Type | Accuracy (%) | | | Precision (%) | | | Recall (%) | | | F1-Score (%) | | |
| --- | --- | --- | --- | --- | --- | --- | --- | --- | --- | --- | --- | --- |
| | CNN-LSTM | GRU | LSTM | CNN-LSTM | GRU | LSTM | CNN-LSTM | GRU | LSTM | CNN-LSTM | GRU | LSTM |
| None | 90 / 94 | 90 / 92 | 91 / 93 | 85 / 93 | 84 / 92 | 87 / 82 | 87 / 91 | 86 / 89 | 87 / 82 | 88 / 92 | 86 / 91 | 87 / 81 |
| FGSM | 49 / 62 | 56 / 64 | 52 / 64 | 43 / 57 | 51 / 64 | 50 / 62 | 42 / 55 | 52 / 59 | 48 / 61 | 41 / 58 | 51 / 65 | 45 / 59 |
| PGD | 29 / 48 | 33 / 47 | 28 / 44 | 28 / 45 | 31 / 46 | 29 / 42 | 34 / 47 | 33 / 47 | 31 / 44 | 29 / 46 | 32 / 48 | 29 / 42 |
| C&W | 18 / 21 | 18 / 21 | 16 / 20 | 18 / 20 | 16 / 20 | 18 / 24 | 20 / 24 | 18 / 20 | 18 / 22 | 17 /20 | 17 /22 | 16 / 21 |

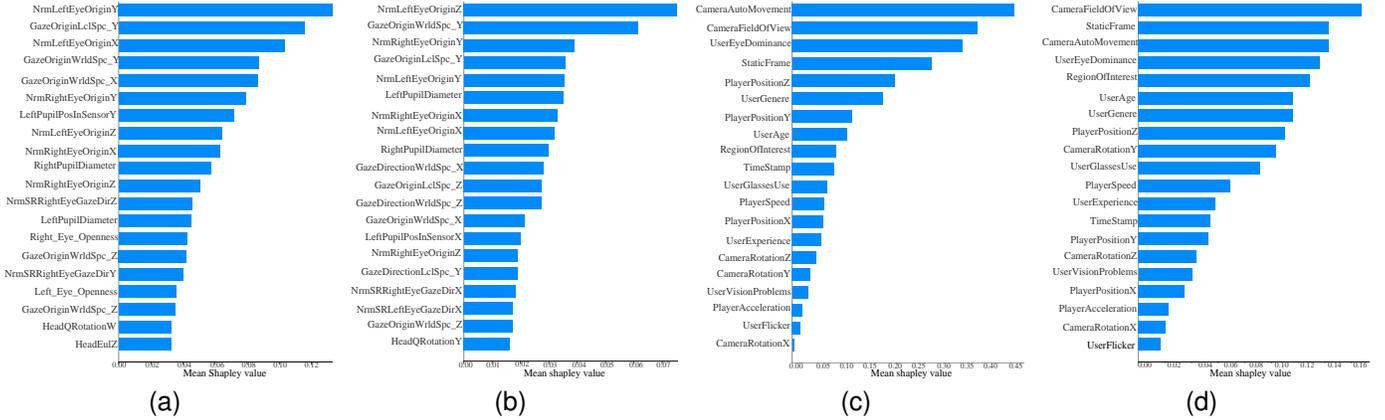

Figure 3: Overall feature importance using global explanation using SHAP on the Simulation 2021 and Gameplay dataset under normal circumstances (**a**) LSTM model for Simulation 2021, (**b**) CNN-LSTM model for Simulation 2021, (**c**) LSTM model for Gameplay, and (**d**) CNN-LSTM model for Gameplay datasets.

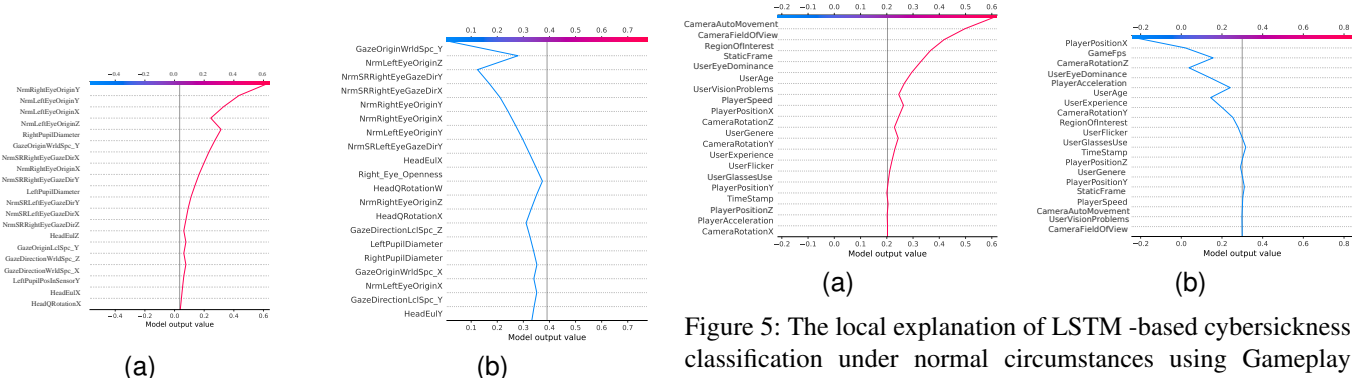

Figure 4: The local explanation of LSTM -based cybersickness classification under normal circumstances using the Simulation 2021 for (**a**) high cybersickness, and (**b**) no cybersickness.

Figure 5: The local explanation of LSTM -based cybersickness classification under normal circumstances using Gameplay dataset for (**a**) high cybersickness, and (**b**) none cybersickness.

are shown in Figure 4. Figure 4(a) and (b) illustrate the high and none cybersickness severity classes. In that figure, the red colored denotes the high cybersickness class, and the blue-colored denotes the low cybersickness class line, showing cybersickness probabilities for that individual outcome. From Figure 4(a), it is observed that the eye tracking feature *NrmRightEyeOriginY* is the most important feature for high cybersickness severity classification, which has the highest probabilities of approximately 0.6. Most of the features except *HeadEulZ*, *HeadEulX*, and *HeadQRotationX* have the highest probabilities of outcome belonging to eye tracking features; thus, an appropriate decision is established for high cybersickness severity classification. For instance, in Figure 4(b), none cybersickness severity classification has a low probability score for eye tracking features, in which *HeadEulY* is the most influential feature with probabilities of 0.3 and *GazeO-*

*riginWrldSpcY* is the least influential feature with probabilities of 0.01. As a result, the correct classification is made for none cybersickness severity class. Likewise, for the gameplay dataset, it is observed that most features contribute to a high cybersickness severity class corresponding to features such as *Player Position Z*, *User glasses use*, *User vision problems*, and *User eye dominance*, and for the none cybersickness severity class, the dominating features are (e.g., *Camera rotation X*, *User vision problem*, etc.), as shown in Figure 5(a) and (b). Such local explanations of feature importance provide insights into the model's classifications and misclassifications and help build trust for the model.

### B. Cybersickness Attack Evaluation

This section presents the results of the impact of white-box and black-box cybersickness attacks on cybersickness models.

*1) Impact of white-box cybersickness attack in cybersickness DL models:* To analyze the impact of cybersickness attacks on the VR cybersickness prediction model, we craft the adversarial examples using FGSM, PGD, and C&W attacks as discussed in Section IV-3, and apply them to three DL models. Figures 6(a), 6(b), and 6(c) show adversarial examples of perturbed data crafted for the LSTM model on the Simulation 2021 dataset using FGSM (with $\varepsilon = 0.1$), PGD with ($\varepsilon = 0.1$, $\alpha = 0.01$, and $I = 20$), and C&W (with max iterations = 1000), respectively. Similarly, Figures 6(d), 6(e), and 6(f) show adversarial examples crafted with these same attacks and models on the Gameplay dataset. From Figure 6, it is observed that the PGD and C&W attacks each generate adversarial examples that more closely match the original data. Such stealthy attacks often fall within the range of the normal VR eye-tracking and head-tracking sensor data, as well as the normal gameplay data. Hence, they are hard to detect using the common attack detection mechanisms, which can potentially cause VR cybersickness by triggering unexpected mitigation techniques. Furthermore, we compute the Pearson correlation coefficient (PCC) [70] to explore the relationship between the benign data and adversarial examples generated by cybersickness attacks for the LSTM model for the Simulation 2021 and Gameplay datasets. For the Simulation 2021 dataset, the PCC values are 0.26, 0.24, and 0.17 for the FGSM, PGD, and C&W attacks. On the other hand, for the Gameplay dataset, the PCC values are 0.21, 0.15, and 0.13 for the FGSM, PGD, and C&W attacks.

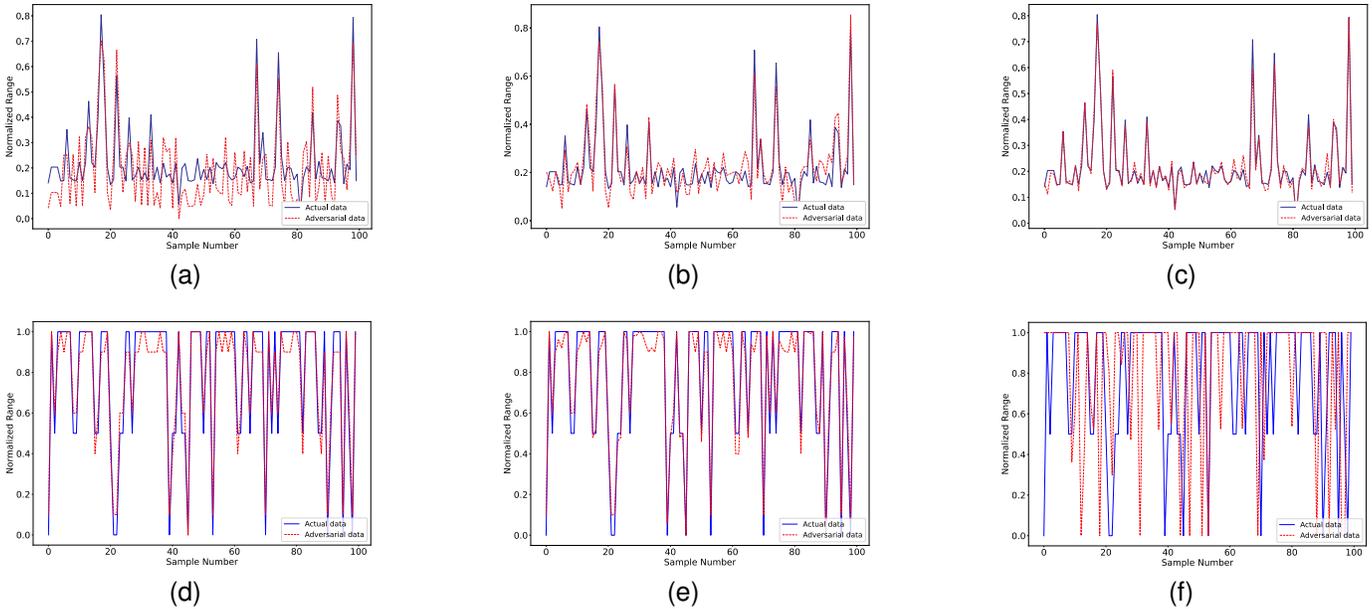

Figure 6: A visualization of normal data vs. cybersickness attacks generated data for the LSTM model, where the *solid* line represents the normal data and the *dotted* line represents the perturbed data. (**a**) and (**d**) show the attack signatures of the FGSM attack. (**b**) and (**e**) show the attack signatures of the PGD attack, and (**c**) and (**f**) show the attack signatures of the C&W attack on the Simulation 2021 and Gameplay datasets, respectively.

From Table II, we observe that the FGSM attack (with $\varepsilon = 0.1$) decreases the accuracy of the LSTM model by approximately 1.75× and 1.45× on the Simulation 2021 and Gameplay dataset, respectively, compared to the accuracies of the model without the attack. Moreover, the PGD attack (with $\varepsilon = 0.1$, $\alpha = 0.01$, and $I = 20$) and C&W attack (with max iterations=1000) decrease the accuracies of the DL models to an even larger extent. For example, the C&W attack causes a 5.69× and 4.65× decrease in accuracy for the LSTM model on the Simulation 2021 and Gameplay datasets, respectively, compared to the accuracies without the attacks. The decreases in accuracy caused by these attacks on the other DL models are extremely similar, showing the attacks' effectiveness on multiple DL model types. It is intriguing that such a small amount of change in the VR measurement data, which is primarily imperceptible to the naked eye, can greatly affect accurate cybersickness detection and trigger unexpected cybersickness mitigation for the VR user, which potentially increases cybersickness for the user.

In addition, to elucidate the impact of FGSM, PGD, and C&W attacks on the cybersickness detection models, we also observe precision, recall, and F1-score for the three DL models, as shown in Table II. Similar to the changes in accuracy, these attacks have an extreme effect on the precision, recall, and F1 scores of the DL models. For instance, on the LSTM model, the C&W attack causes a decrease in precision, recall, and F1-score of 4.8×, 4.84×, and 5.45× on the Simulation 2021 dataset, and 3.45×, 3.72×, and 3.85× on the Gameplay dataset, respectively. In all cases, we observe that both the PGD and C&W attacks have greater effects on the precision, recall, and F1-score of the DL models than the FGSM attack.

*2) Impact of black-box cybersickness attack in cybersickness DL models:* We perform a comprehensive transferability analysis for evaluating the impact of black-box cybersickness attacks (e.g., FGSM, PGD, and C&W attacks) on cybersickness detection models. Specifically, we apply the adversarial examples crafted for a cybersickness detection model to the other cybersickness detection models using the Simulation 2021 and Gameplay datasets. In black box attack scenarios, the attacker has no knowledge about the target model's internal parameters [71], but it can still considerably impact the target model. Our obtained results are presented in Table III. It is observed that the FGSM, PGD, and C&W adversarial examples crafted for the GRU model give lower accuracy when transferred to other DL-based cybersickness models for both Simulation 2021 and Gameplay datasets. Interestingly, when adversarial examples are crafted using FGSM and PGD, they show better transferability by reducing accuracy. For instance, the GRU-based cybersickness detection model has an accuracy of 90% and 92% for Simulation 2021 and Gameplay datasets, respectively (without attack). When we craft adversarial examples for the GRU model using FGSM, PGD, and C&W and transfer them to the LSTM model, we observe that C&W adversarial example decreases the accuracy by 37%, which is almost $1.13\times$, and $1.90\times$ lower for the Simulation 2021 dataset and 51% almost $1.17\times$, and $1.39\times$ lower for the Gameplay dataset when compared to PGD and FGSM attacks, respectively. A similar trend is also observed for CNN-LSTM and LSTM models. Thus, the more complex the cybersickness attack is, the more portable it is across model types.

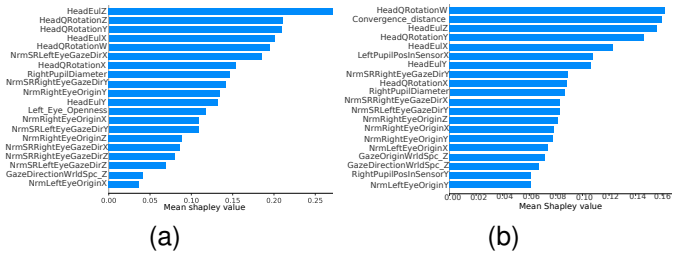

Figure 7: Overall feature importance using SHAP on Simulation 2021 dataset after C&W attack (**a**) LSTM, and (**b**) CNN-LSTM model.

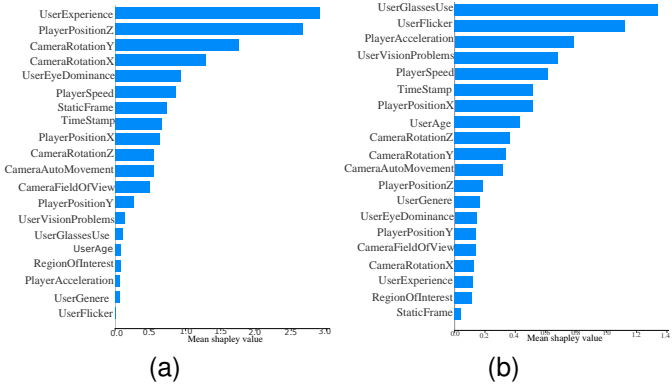

Figure 8: Overall feature importance using SHAP on Gameplay dataset after C&W attack (**a**) LSTM, and (**b**) CNN-LSTM model.

Table III: Transferability of FGSM, PGD, and C&W attacks. The notation X / Y / Z represents model accuracy after FGSM / PGD / C&W attacks.

| Generating Models | Accuracy (%) | | | | | |
|---|---|---|---|---|---|---|
| | Simulation 2021 Dataset | | | Gameplay Dataset | | |
| | CNN-LSTM | GRU | LSTM | CNN-LSTM | GRU | LSTM |
| CNN-LSTM | – | 80 / 54 / 43 | 89 / 48 / 39 | – | 61 / 52 / 48 | 59 / 62 / 71 |
| GRU | 85 / 53 / 49 | – | 80 / 42 / 37 | 84 / 67 / 50 | – | 71 / 60 / 51 |
| LSTM | 84 / 64 / 49 | 79 / 61 / 47 | – | 79 / 73 / 62 | 71 / 64 / 54 | – |

### C. Cybersickness Attack Defense Evaluation

This section presents the results of XAI signature generation and cybersickness attack detection using these XAI signatures.

*1) Attack signature generation:* This section applies SHAP-based global and local explanation methods to explain the DL model's cybersickness classifications under cybersickness attack conditions. Figure 7 illustrates the overall feature importance for cybersickness severity classification using the LSTM and CNN-LSTM models for adversarial examples generated by the C&W attack on the Simulation 2021 dataset. From Figure 7(a) and (b), we observe that features such as *HeadEulZ*, *Convergence distance*, *HeadQRotationW*, etc., are the most influential features in cybersickness classification after the cybersickness attack. This shows that the effect of a cybersickness attack is greatly impacted by feature importance, thus significantly identifying the wrong feature (e.g., *Convergence distance* and *HeadQRotation*) in the model outcome. Thus, we observe that head-tracking data such as *HeadQRotationW*, *HeadEulZ*, etc., is much more influential for the cybersickness severity classification of CNN-LSTM and LSTM models. Likewise, from Figure 8(a) and (b), we observe that the features such as *User experience*, *User glasses use*, *User flicker*, *Camera rotation Y*, *Player acceleration*, etc., are the most important features in cybersickness classification for the LSTM and CNN-LSTM models using the Gameplay dataset under C&W attacks. Due to the attacks, the most important features have drastically changed, similar to the Simulation 2021 dataset.

The local explanation of the classified cybersickness for the LSTM model under cybersickness attack (C&W-generated adversarial examples) using the Simulation 2021 and Gameplay datasets is shown in Figure 9. Due to the cybersickness attack, there are wild spikes in the graph, showing that almost all of the features shown are making significant changes to the model's decision, which is not true for normal conditions (no attack). Moreover, the order of importance of these features has also changed due to the attacks. For instance, from Figure 9(a), we observe that, like with the global explanations, features such as *HeadEulX*, *HeadQRotationZ*, *Convergence distance* etc., have the most influence in making the decision regarding high cybersickness classification for the LSTM model, which indicates that most of the features that contribute to the positive impact are the least important features to trigger cybersickness. On the other hand, for the LSTM model on the Simulation 2021 dataset under normal circumstances (no attack condition), *NrmLeftEyeOriginY* is the second most influential feature for the high cybersickness severity classification. However, as shown in Figure 9(b), it is not listed as a dominant feature after the C&W attack. Likewise, Figure 9(a) illustrates the none cybersickness results for the LSTM model after the C&W attack. We observe that head tracking features such as *HeadEulZ*, *HeadQRotationY* are the most influential features for none cybersickness class classification with a positive feature importance score. If we observe Figure 9 (a) and (b) very closely, for both of these predictions, head tracking features are the most predictive features for high cybersickness and none cybersickness classification severity classification. This is also the case for the Gameplay dataset. For instance, after the C&W attack, *User experience* (which had a minimal effect under normal circumstances (without attack)) is the most influential feature on the high cybersickness severity classification, and *Time Stamp* is the most influential feature on the none cybersickness severity classification for the LSTM model on the Gameplay dataset, as shown in Figure 9(c) and (d). This local explanation of feature importance makes the difference between adversarial and normal examples significantly easier to detect. Then, we use these explanations denoted as *XAI signatures* as features for our adversarial example detector.

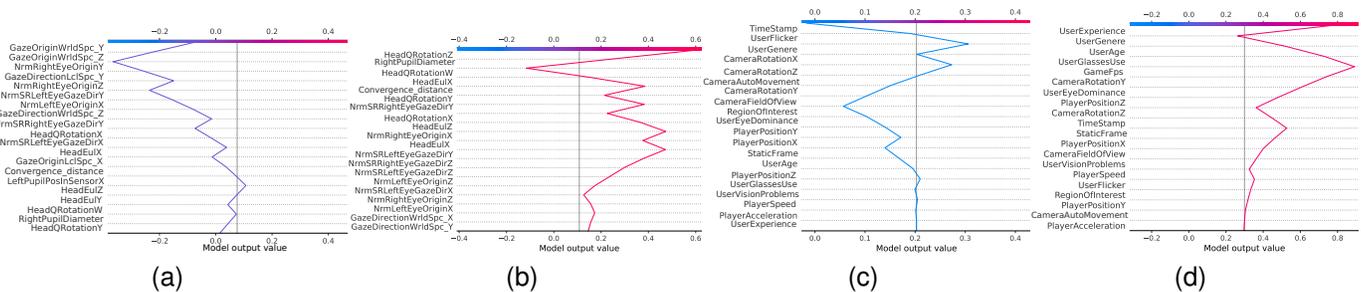

Figure 9: The local explanation after C&W attack of LSTM-based cybersickness classification model (**a**) none, and (**b**) high cybersickness using Simulation 2021 dataset and (**c**) none, and (**d**) high cybersickness using Gameplay dataset.

*2) Cybersickness attacks detection using XAI signatures:* From the previous sections, we observe that the SHAP-based local explanation method provides insight into how the feature importance changes under cybersickness attacks that can significantly impact accurate cybersickness prediction, eventually leading to unexpected mitigation. In this section, we apply our proposed XAI-guided cybersickness attack detection method. For detecting the adversarial examples, we train two ML models, namely RF and XGB, and one DL model, FFNN to classify the normal and adversarial examples using the XAI signatures of both normal (N) and adversarial (A) examples. Table IV summarizes the accuracy and F1-score of the cybersickness attack detection using RF, XGB, and FFNN models for both Simulation 2021 and Gameplay datasets. We show that our attack detection methods can differentiate adversarial and normal examples with high accuracy. For instance, for the CNN-LSTM model, the XGB model achieves an accuracy of 90% and 94% using the Simulation 2021 and Gameplay datasets, respectively, which is $1.02\times$ and $1.04\times$ and $1.07\times$ and $1.02\times$ higher than FFNN and RF models. Indeed, the overall performance of the XGB model is slightly better than the RF and FFNN models in terms of F1-score for normal and adversarial example identification. Surprisingly, we observe simple ML models perform better than the FFNN model. A similar trend is also observed for the GRU and LSTM models' cybersickness attack detection.

## VIII. TESTBED IMPLEMENTATION AND EVALUATION

This section explains the details of our testbed implementation and presents the results of our implemented testbed in terms of cybersickness attack and defense.

### A. Testbed Implementation

Our VR testbed comprises two essential parts: (1) a local VR environment and (2) a cloud-based environment. Specifically, the local VR environment runs the VR simulation and cybersickness mitigation engine. On the other hand, the cloud-based environment includes a deployed DL model, XAI-guided attack signature generation, and a cybersickness attack detector for detecting cybersickness attacks. It is important to note that we developed a custom VR testbed to meet the specific requirements of our study, i.e., integrating real-time DL-based cybersickness detection with adaptive mitigation techniques and providing precise control over experimental parameters. The custom testbed also allow us to tailor the user study environment to the unique goals of our research, such as testing adversarial robustness in cybersickness prediction models. We assume the adversary has access to local VR and cloud environments. The adversary aims to fool the cybersickness detection model to misclassify the user's cybersickness prediction and trigger unexpected mitigation, hampering the UIX as shown in Figure 10. For instance, as shown in Figure 10a, an unexpected mitigation technique is triggered due to the cybersickness attack; e.g., the cybersickness detection model predicts the user's cybersickness level as severe, whereas the participant is not cybersick. On the other hand, while the same user used our proposed XAI-guided cybersickness attack detection method, the correct mitigation was triggered by successfully detecting the cybersickness attack, as shown in Figure 10b. The details of the testbed are as follows.

Table IV: Accuracy (Acc.) and f1-score (F1-S) of cybersickness attack detection ("XAI signatures") of the cybersickness models on (Simulation 2021 dataset/Gameplay dataset).

| Model | CNN-LSTM | | | GRU | | | LSTM | | |
|---|---|---|---|---|---|---|---|---|---|
| | Acc. | F1-s | | Acc. | F1-S | | Acc. | F1-S | |
| | | N | A | | N | A | | N | A |
| FFNN | 88 / 88 | 88 / 87 | 88 / 88 | 78 / 90 | 77 / 90 | 79 / 90 | 86 / 87 | 86 / 87 | 86 / 87 |
| RF | 86 / 92 | 85 / 92 | 86 / 92 | 78 / 94 | 76 / 93 | 80 / 94 | 84 / 90 | 83 / 90 | 86 / 91 |
| XGB | 90 / 94 | 90 / 94 | 89 / 94 | 82 / 95 | 81 / 95 | 82 / 95 | 87 / 92 | 87 / 92 | 88 / 92 |

### LOCAL VR ENVIRONMENT

**Apparatus:** Our proposed environment is implemented as a Unity project compatible with an eye and head-tracking-enabled HTC-Vive Pro Eye VR headset. The headset offers a display resolution of 1440 × 1600 per eye and operates at a refresh rate of 90Hz. The headset provided a wide field of view, spanning 110 degrees. The virtual environment was crafted using Unity 3D. To capture eye-tracking and head-tracking data, we utilize the HTC SRanipal SDK and Tobii HTC Vive Devkit API [72]. Our testbed uses a roller coaster simulation, widely used in SOTA cybersickness detection and reduction research [24], [34], [73], further enhancing its relevance and applicability. We ran our experiments using two

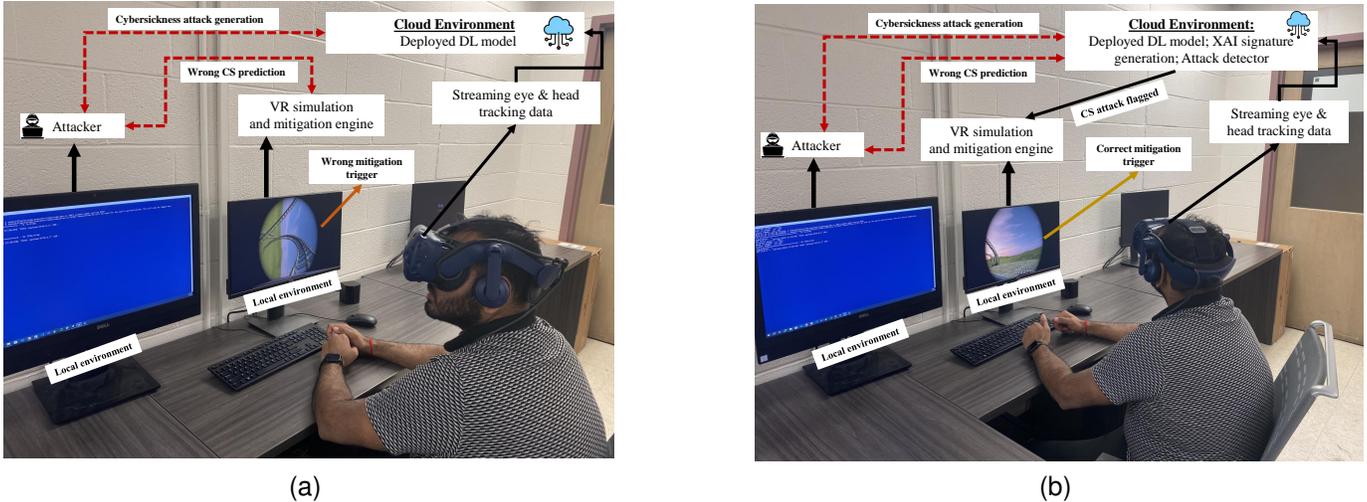

Figure 10: Participant immersed in our proposed testbed (**a**) under cybersickness attack condition, and (**b**) Under the cybersickness attack detection phase, in which the XAI-guided method detects cybersickness attacks and alerts the VR user, thus triggering the correct mitigation.

PCs with an Intel Core i9 processor and a 32GB RAM option with an NVIDIA GeForce RTX 3080 Ti GPU.

**Cybersickness Mitigation:** We use GingerVR [33] software package to develop the mitigation engine for the Unity game engine. In this work, we develop the mitigation library using three popular cybersickness mitigation techniques: dynamic Gaussian blurring, dynamic field-of-view (FOV), and foveated depth-of-field (FOF) blur. Figure 11(a), Figure 11(b), Figure 11(c), and Figure 11(d) show the example of the output of the no mitigation applied, dynamic FOV reduction, dynamic Gaussian blurring, and foveated DOF effect in cybersickness detection for the VR roller coaster simulation.

### CLOUD ENVIRONMENT

**DL model deployment and Inference phase:** After training and validation, the DL-enabled cybersickness detection model is converted to Keras format [74] and deployed on an HTC Vive VR headset. In the inference phase, the pre-trained deployed DL model predicts the severity of the cybersickness (none, low, medium, and high) using streaming VR simulation eye and head-tracking data. The DL model predictions are then forwarded to the mitigation engine. Suppose the user's predicted cybersickness is flagged as *high, medium*, or *low*. In that case, the mitigation engine will be triggered to activate the correct mitigation technique. If the user's level of cybersickness is flagged as no cybersickness denoted as *none*, no mitigation techniques will be triggered for updating the frame.

**Cybersickness attack and XAI signature generation:** We use the CleverHans [69] library to implement the proposed cybersickness attack. We used SHAP-based post-hoc explanation methods to explain the DL models and calculate the feature importance score for XAI signature generation. For this, we used the SHAP [65] library.

**Cybersickness attack detection:** After training and validation, the ML/DL-enabled cybersickness attack detector is deployed in a cloud environment to detect the cybersickness attack in real-time.

### B. Testbed Evaluation

This section presents the detailed results of our proposed testbed evaluation, which is necessary for the user study. Specifically, we explain the deployed DL model, the cybersickness attack generation, and the XAI-guided signature generation in the testbed in detail. We used the trained LSTM-based cybersickness model (since LSTM performs the best in terms of cybersickness detection as shown in Table II), FGSM, PGD, and C&W attacks, SHAP explanation method, and XGB-based cybersickness attack detector for this testbed to conduct the user study. Moreover, we only used the Simulation 2021 dataset for training the LSTM model and deploying it in the HTC Vive Pro headset to conduct the experiment. This is because we are interested in detecting cybersickness attacks on the integrated sensor (e.g., eye-tracking and head-tracking) during the VR simulation. The total size of the deployed LSTM model on the HTC Vive Pro headset is 280,917 KB. In addition, the training time for the deployed DL model is 54,000 seconds. In the VR simulation, the deployed LSTM model requires only 0.0018 seconds to predict the cybersickness severity per frame. During the experiment, we inject the cybersickness attack after 1 minute of VR simulation, and the attack lasts for 2 minutes. In this study, we restrict ourselves to only white-box attack scenarios in which the adversary has complete knowledge of the cybersickness detection models, which helps craft powerful adversarial examples to test our attack detection method. Figure 12 shows the actual and predicted cybersickness level for both normal and adversarial conditions for a 180 seconds time window in the proposed testbed. It is observed that in the cybersickness attack condition, the prediction model predicts the wrong cybersickness class. For instance, in 150 seconds, the model predicts a cybersickness class as *none* under the cybersickness attack condition, while the actual cybersickness class was *high* in normal conditions (no attack). Therefore, no mitigation will be triggered, and the user feels severe cybersickness.

Next, let us evaluate the performance of the proposed cybersickness attack detector, where each cybersickness level prediction outcome (none, low, medium, high) passes through the detector. To do this, we first generate the XAI signature using SHAP local explanation to detect a cybersickness attack. After generating the attack signatures, the cybersickness attack detector detects the cybersickness attack and sends an alert to the user. Figure 13 shows different mitigations triggered for different conditions (e.g., no attack vs. cybersickness attack vs. our proposed XAI-guided cybersickness attack detection method) for a 180 seconds time window in the proposed testbed. It is observed that in the cybersickness attack condition, no attack is detected, and the wrong mitigation is triggered; thus, the user feels cybersickness. In contrast, our proposed XAI-guided cybersickness attack detector detects the attack successfully; thus, the correct mitigation technique is usually triggered, effectively reducing cybersickness and increasing UIX. For instance, in 150 seconds, the foveated depth-of-field (FOVR) mitigation technique is triggered for both no attack and our proposed method condition. However, no mitigation technique is triggered in the cybersickness attack condition, which shows our proposed method successfully detects the attack and thus triggers the proper mitigation method to reduce the cybersickness.

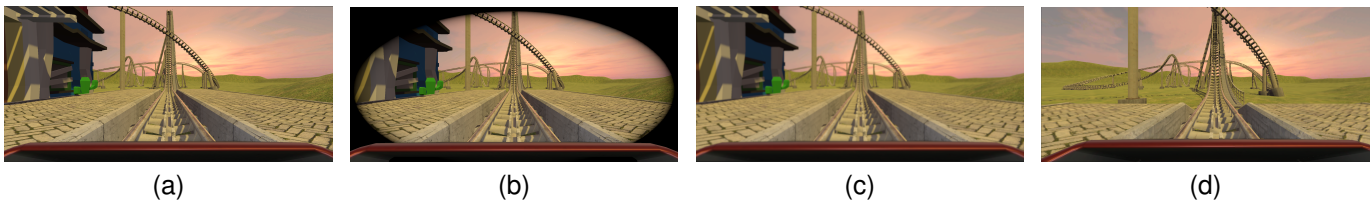

(a)  (b)  (c)  (d)

Figure 11: The user's perspective within the developed testbed (roller coaster simulation) when mitigation techniques are applied: (**a**) no mitigation applied, (**b**) dynamic FOV reduction, (**c**) dynamic Gaussian blurring, and (**d**) foveated DOF for different conditions.

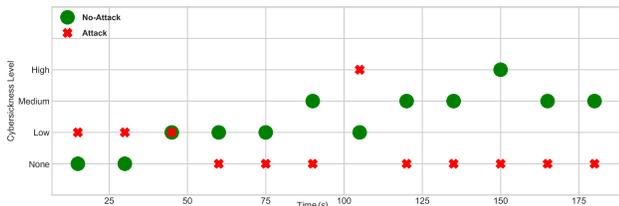

Figure 12: Cybersickness prediction under no attack vs. cybersickness attack conditions in the testbed.

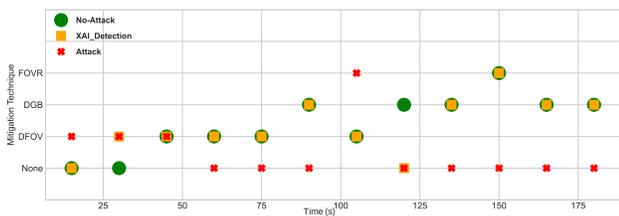

Figure 13: Cybersickness mitigation trigger under different conditions: no-attack condition vs. cybersickness attack condition vs. XAI-guided attack detection scheme in the testbed.

## IX. USER STUDY DESIGN AND EVALUATION

This section presents the user study design and results of the user study evaluation to demonstrate the effectiveness of our proposed cybersickness attack detection framework.

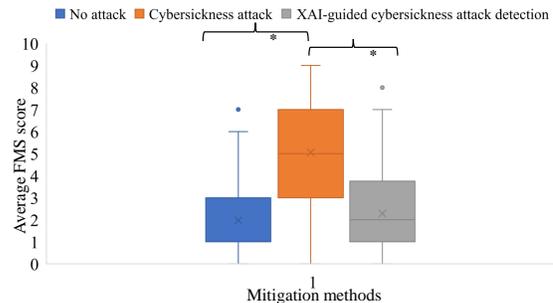

Figure 14: Average FMS score for different conditions for 30 participants.

Table V: The Wilcoxon rank pair test between the different conditions of the FMS score test.

| Condition | Z-value | P-value | W-value |
|---|---|---|---|
| No attack vs. cybersickness attack | 5.41 | **.00001** | 124 |
| No attack vs. XAI-guided | 1.16 | .250 | 475 |
| Cybersickness attack vs. XAI-guided | 5.29 | **.00001** | 141 |

### A. Participants, Virtual Environment, Task, and Procedure

This user study involves 30 volunteer participants who received no reward (18 males and 12 females) aged from 22 to 38 years (mean age 26.9, standard deviation (SD) 6.45). Before starting the experiment, we briefly introduced the procedures and informed participants about the requirements. 23 participants said they had previously used VR equipment, while 7 had no VR experience. We emphasized verbally reporting their discomfort on an FMS score of 0 to 10 whenever a pre-recorded voice prompt was played at each 30 seconds interval during the VR simulations [24]. A rating of 0 means no significant change in comfort level compared to the resting baseline, and 10 means significant discomfort compared to resting conditions. At the beginning of each trial, participants had a simple practice session to help them become familiar with how to use the controllers for different locomotion techniques. After completing the trial session, the roller coaster VR simulations were played and lasted 10 minutes long unless the participants decided to quit earlier. It is important to note that the eye and head positions were calibrated before starting the simulation. Upon satisfactory calibration, participants were immersed in a VR trial session. After completing the VR simulation with and without attack conditions, participants were asked to complete an SSQ questionnaire [75] and then rest. We allowed the participants to rest as much as they wanted until they thought they could continue with the next session without any negative feelings. The experiment was over when they completed applying all methods, such as the baseline condition (no attack), the cybersickness attack condition, and the XAI-guided cybersickness attack detection condition. To validate our proposed approach, in which mitigation was triggered/not triggered, we generated the following three conditions for the experiment: 1) At first, we ran the simulation as a baseline condition in which no attacks were applied, 2) Then, as discussed in section IV, we applied the cybersickness attack in which the suitable mitigation technique is either triggered/not triggered, and 3) Finally, as discussed in Section V, we applied the XAI-guided cybersickness attack detection technique in which the mitigation technique is triggered after the correct cybersickness attack is flagged. Let us recall that our proposed XAI-guided cybersickness attack detector sends an alert to the user once the cybersickness attack is detected, and then, the adversarial sample is discarded. Only the samples classified as normal (without attack) are forwarded to the mitigation engine to trigger the proper mitigation technique. It is worth mentioning that all the conditions were run in counter-balanced order. However, participants could stop at any point for any reason (e.g., if they felt highly uncomfortable) during the experiment. Furthermore, before participants left the room, we reminded them that they could reach out to us if they felt severe discomfort or experienced any negative impact from the study. No one mentioned it, nor did we observe any concern or severe discomfort.

**Hypotheses:** The primary focus of this research is to identify and understand the impact of cybersickness attacks that can severely hamper UIX by triggering unexpected mitigation. Therefore, to detect cybersickness attacks, the following two hypotheses (**H**) are formulated to form the basis of an experiment and evaluate the proposed XAI-guided cybersickness attack detection method.

1) **H1:** *Cybersickness attacks successfully induce cybersickness by fooling the cybersickness prediction model outcome and triggering unexpected mitigation techniques..*

2) **H2:** *XAI-guided cybersickness attack detection method successfully flagged the attack, sent an alert to the user, and thus prevent triggering incorrect/unnecessary mitigation technique (as a consequence of cybersickness attack) reducing participant's UIX.*

### B. User Study Evaluation

This section presents the detailed results of the user study for evaluating the effects of cybersickness attacks and XAI-guided attack detection. We analyze our experimental data to study users' reactions when encountering cybersickness attacks during VR experience. Then, we analyzed the post-task user study experimental data after encountering cybersickness attacks in the VR experience. The details are below.

*1) Experiment phase:* The mean, standard deviation, and 95% confidence interval of each condition's measured FMS score are shown in Table VI. The obtained results show that the FMS score is highest in cybersickness attack conditions, ranging between $2.12 - 2.69$ in the 95% confidence intervals, meaning that cybersickness attacks increase user cybersickness, while the measured FMS score is relatively lower in the no attack and XAI-guided cybersickness attack detection conditions. To test the significance of these differences in means, we first conduct a Shapiro-Wilk test [76], after which we determine that the FMS score per 30 seconds of data is not normally distributed. Therefore, we conduct a non-parametric test, namely the Friedman test [76], on these experimental data instead of ANOVA [76]. We set our significance level $Alpha(\alpha) = 0.05$. The Friedman test helps to investigate if there are significant differences in cybersickness between the conditions: the baseline (without attack), with attack, and our XAI-guided attack detection ($\chi^2$ = 47.425, P = .001). This Friedman test, specifically the measured P-value, provided significant evidence that the obtained FMS score per 30 seconds data, meaning that each condition has significance with their results in FMS measurement. For post hoc comparisons, we conduct Wilcoxon signed-rank tests [77] with Bonferroni correction [78] for each condition, as shown in Table V. A significant difference is found in the attack condition vs. non-attack vs. XAI-guided attack detection scheme by comparing the measured P-value (P < ".05"). The results indicate that the cybersickness attack has a significant difference, in which the experimental environment caused a significant increase in the FMS scores. For instance, the P-value for the Wilcoxon paired test in **no attack vs. cybersickness attack**, and **cybersickness attack vs. XAI-guided cybersickness attack detection** is (P < ".05"). However, we do not find any significant difference when comparing the **no attack vs. XAI-guided cybersickness attack detection** condition (P > ".05"). Figure 14 shows the average FMS score of the different applied conditions in cybersickness reduction for different users. This box plot shows that the cybersickness attack increases cybersickness by significantly increasing users' FMS scores. Furthermore, Bonferroni correction is used for pairwise comparison if a significant difference is found in the Wilcoxon paired test. The Bonferroni test shows a significant difference for the **no attack vs. cybersickness attack**, and **cybersickness attack vs. XAI-guided cybersickness attack detection** conditions (P < ".05"). However, we do not find a significant difference between no attack and XAI-guided cybersickness attack detection conditions (P > ".05"). Furthermore, to show the effectiveness of our proposed method, we calculate the effect size for each condition. We find a relatively medium effect for **no attack vs. cybersickness attack** and **cybersickness attack vs. XAI-guided cybersickness attack detection** conditions are .805 and .793, respectively. However, we find a minimal effect between no attack and XAI-guided cybersickness attack detection conditions (e.g., .087). The results show that a cybersickness attack severely induces cybersickness by increasing the user's FMS score. However, our proposed XAI-guided cybersickness attack detection method detects the cybersickness attack, which helps trigger the proper mitigation method, which reduces cybersickness in VR.

Table VI: The mean, standard deviation, and 95% confidence intervals of the FMS score for each condition.

| Condition | Mean (SD) | 95% Confidence Interval |
|---|---|---|
| No attack | 1.89 (1.98) | [1.475-2.305] |
| CS attack | 4.89 (2.58) | [4.260-5.535] |
| XAI-guided CS attack detection | 1.91 (2.06) | [1.389-2.391] |

*2) Post-task interview Phase:* We also provide additional user study results, demonstrating how well the proposed XAI-guided cybersickness attack detection framework makes the VR experience smooth and comfortable by detecting cybersickness attacks and initiating the proper mitigation technique. Note that after detecting the cybersickness attack, the adversarial sample is discarded, and only the normal sample is forwarded to the mitigation phase to trigger the proper mitigation technique. For this, during the user study, we asked users to rate their experiences using the XAI-guided cybersickness attack detection framework on a 1 to 5 in the Likert scale [79] compared to cybersickness attack conditions in which the unexpected mitigation is triggered. More specifically, we asked users how much they agreed with the following statements (S): 1) *The framework helped detect cybersickness attacks, trigger the proper mitigation and complete the tasks ( e.g., staying engaged in the VR simulation)*, 2) *Based on my experience and using the framework so far, the decision regarding cybersickness and triggering the mitigation is correct under cybersickness attack condition*, 3) *To what extent the proposed framework enhance your overall VR experience*, 4) *Is this framework built your trust towards the VR experience?*, 5) *How comfortable you feel using the framework for cybersickness attack detection and mitigation technique trigger throughout your VR session?*, 6) *What concerns do you have if you don't feel comfortable completing tasks in VR using this framework?* 7) *Do you have any concerns about adopting this technology in your daily life?*, and 8) *Would you like to recommend this framework to others who use VR?* To evaluate the rated users' experiences, we compute the mean and standard deviation of the Likert score for these queries for each user. It is important to note that we compute this between the cybersickness attack condition and the XAI-guided cybersickness attack detection condition. The user feedback results are provided in Figure 15. We observe that most users, approximately 93%, agreed with all statements to use our proposed XAI-guided cybersickness attack detection framework for detecting cybersickness attacks and triggering the proper mitigation technique to reduce cybersickness and ensure a smooth and comfortable VR experience. At the same time, most users agree not to use the traditional approach (without XAI-guided cybersickness attack detection) due to their ineffective UIX and less engagement in the VR environment. Moreover, the user had an uncomfortable VR experience while engaging in the traditional approach, which sometimes forced them to quit the VR simulation.

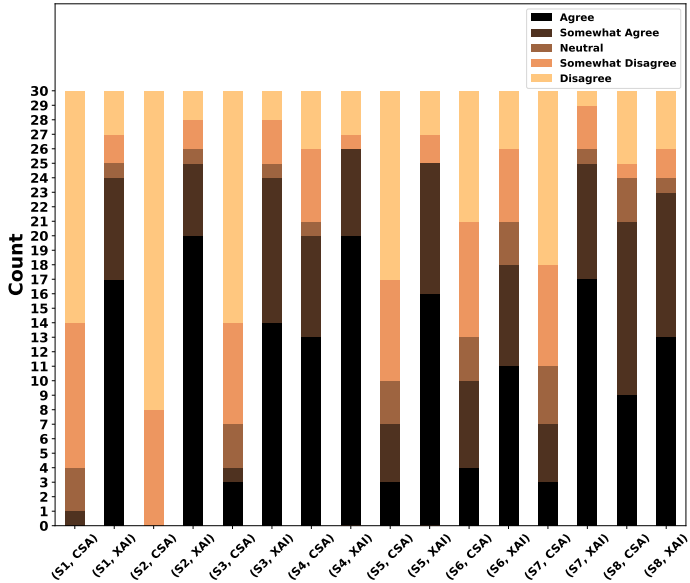

Figure 15: User feedback in Likert scale between our proposed XAI-guided cybersickness attack detection method denoted as **(XAI)** framework and without XAI-guided cybersickness attack detector framework denoted as **(CSA)** for evaluating the impact of cybersickness attack on UIX.

## X. DISCUSSIONS

This section briefly discusses the results obtained using the XAI-guided cybersickness attack detection method. Technical evaluation and user-based studies have demonstrated that cybersickness attacks dramatically hinder the UIX by successfully fooling the cybersickness prediction model outcome and triggering unexpected cybersickness mitigation techniques, which supports our **H1**. Most participants reported feeling less discomfort and ensuring a smooth and comfortable VR experience after experiencing our framework. Our experimental results show that cybersickness attacks can successfully fool the cybersickness detection system. For instance, the C&W attack causes a $5.69\times$ and $4.65\times$ decrease in accuracy for the LSTM-based cybersickness detection model on the Simulation 2021 and Gameplay datasets, respectively, compared to the accuracies without the cybersickness attacks successfully stops the triggering of cybersickness mitigation and harming UIX. However, our XGB-based cybersickness attack detector successfully identified XAI-generated *attack signatures* with 90% (Simulation 2021 dataset) and 94% (Gameplay dataset) accuracy. User feedback confirmed a preference for our XAI-guided framework over conventional methods, which are vulnerable to attacks and disrupt immersion, highlighting its effectiveness in maintaining engagement and comfort in VR.

Since no previous work applies adversarial perturbation to fool cybersickness detection model and XAI-guided attack detection methods, our results are not directly comparable with prior works in this area [1], [11]–[13], [39], [79]. However, we compare our results regarding users' smooth and comfortable VR experience while they used our proposed method. Prior work focused mainly on developing DL-enabled accurate cybersickness prediction models and mitigation techniques rather than looking into the vulnerability of these models [21], [26], [27], [34]. In contrast, our work manipulates the vulnerability of DL models for cybersickness detection models to propose new cybersickness attacks and use XAI-based methods to detect them. For instance, after experiencing a cybersickness attack condition, one participant reported that he felt severe discomfort and was disoriented because the mitigation was not triggered correctly. And at last, he left the simulation due to his severe discomfort. On the other hand, the same participant reported that he felt no cybersickness for the whole experiment and enjoyed a smooth and comfortable VR experience while using our proposed framework. This is because the attacker could not trigger incorrect (when the user is cybersick) and unnecessary mitigation (when the user is not cybersick) by fooling the cybersickness detection model.

Instead, when an attack is detected, the user receives an alert, which increases users' trust in the VR environment. These results support **H2** in which participants had a better UIX using the proposed XAI-guided cybersickness attack detector. Another participant realized that unexpected mitigations were triggered under cybersickness attack conditions, even though he did not feel cybersickness during his session. However, the same participant reported that while he used our proposed framework, he did not witness such an unexpected mitigation-triggering experience. On the other hand, two participants reported that defensive strategies sometimes helped them avoid cybersickness attacks, but not every time, which is true since there are a limited number of cases when our proposed cybersickness attack detection fails to detect the cybersickness attack (see Figure 13 in section VIII-B).

## XI. Limitation and Future Works

Our XAI-guided cybersickness attack detection framework has some limitations. While the CNN-LSTM and XGB models achieve 90% and 94% accuracy in detecting attacks on the Simulation 2021 and Gameplay datasets, respectively, 6%-10% of cases still result in detection failures. This can lead to unnecessary or incorrect mitigation triggers, as reported by some participants during the user study. Additionally, our detector may be unable to detect the new unseen attacks, as it is trained only on FGSM, PGD, and C&W attacks, leaving it vulnerable to novel adversarial manipulations. Future work will expand the framework to add more attacks and detect more advanced attacks in real-time. Moreover, our study also had a limited sample size, focusing on a single VR experience (roller coaster simulation) with imbalanced gender representation. Since cybersickness might affect people based on their unique characteristics, tasks, and VR environment [29], [49]. Also, our findings are based on a specific hardware-software setup, which may not be generalized to other setups or future technologies. Future research should explore diverse VR scenarios, extended exposure times, interactive tasks, and a more balanced demographic to enhance the framework's applicability. Additionally, exploring more VR simulations, attacks, and mitigation techniques will further strengthen the robustness of our approach.

## XII. Conclusion

In this work, we proposed an XAI-guided robust cybersickness attack detection and mitigation framework by leveraging explainable AI for cybersickness attack detection and triggering the proper mitigation method to ensure a smooth and comfortable VR experience. Specifically, we showed that LSTM, CNN-LSTM, and GRU cybersickness detection models could be fooled by launching the FGSM, PGD, and C&W attacks, and unnecessary cybersickness mitigation can be triggered. For instance, our experimental results showed that C&W attack causes a $5.69\times$ and $4.65\times$ decrease in accuracy for the LSTM-based cybersickness detection model on the Simulation 2021 and Gameplay datasets, respectively, compared to the accuracy without the cybersickness attacks. We also proposed an ML and DL-based adversarial example detection method to detect cybersickness attacks by differentiating XAI-generated *attack signatures* from the cybersickness detection models. The proposed detection method successfully detected cybersickness attacks with an accuracy of 90% (Simulation 2021 dataset) and 94% (Gameplay dataset). Furthermore, we implemented a testbed on the HTC Vive Pro to validate our proposed attack and defense methods through a mixed-factorial user study. Our user study revealed that the aptness of our proposed XAI-guided cybersickness attack detection method accurately flagged the cybersickness attacks, sent alerts to the users, and triggered the proper mitigation technique, which reduced VR cybersickness with marginal impacts on UIX. In addition, 93% of users agreed that the proposed framework is highly effective for a smooth, safer, and more comfortable VR experience. We believe the proposed XAI-guided cybersickness attack detection framework will be helpful for other researchers, developers, testers, and VR users for developing, testing, and using cybersickness detection and mitigation methods on actual consumer-level VR headsets.

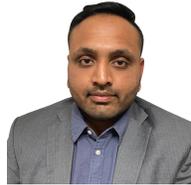

**Ripan Kumar Kundu** received his B.Sc.degree in electrical and electronic engineering from the American International University Bangladesh (AIUB), Dhaka, Bangladesh, in 2016, and his M.S. degree in electrical engineering from the University of Rostock, Rostock, Germany, in 2022. He is currently pursuing his Ph.D. degree in Computer engineering and working as a graduate research assistant in the DCPS Laboratory at the University of Missouri-Columbia. His research interests include trustworthy AR/VR, and Explainable & Fair AI/ML.

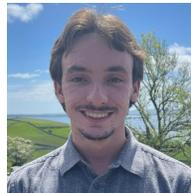

**Matthew Denton** is currently working toward a BS degree in computer science with an emphasis in Artificial Intelligence and a minor in mathematics with Washington University, Saint Louis. His research interests include virtual reality and Artificial Intelligence.

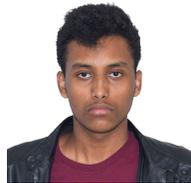

**Genova Mongalo** is currently working toward the BS degree in computer science with an emphasis in cyber security and a minor in mathematics with University of Missouri, Kansas City. His research interests include virtual reality, and cyber security.

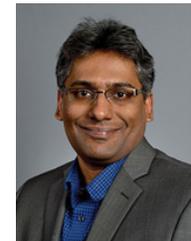

**Prasad Calyam** received his MS and PhD degrees from the Department of Electrical and Computer Engineering at The Ohio State University in 2002 and 2007. He is currently the Greg L. Gilliom Professor of Cybersecurity in the Department of Electrical Engineering and Computer Science at University of Missouri-Columbia. Previously, he was a Research Director at the Ohio Supercomputer Center. His current research interests include distributed and cloud computing, computer networking, and cybersecurity. He is a Senior Member of IEEE.

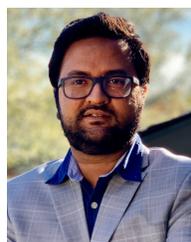

**Khaza Anuarul Hoque** is an Assistant Professor in the Department of Electrical Engineering and Computer Science at the University of Missouri-Columbia (MU), where he directs the Dependable Cyber-Physical Systems (DCPS) Laboratory. He received his M.Sc. and Ph.D. from the Department of Electrical and Computer Engineering at Concordia University, Montreal, Canada, in 2011 and 2016. His research interests include cyber-physical and embedded Systems, formal methods, cybersecurity, and safe AI/ML. Before joining MU, he was an FRQNT postdoctoral fellow at the University of Oxford, UK. Dr. Hoque has received several awards and distinctions, including the 2023 IEEE/ACM MEMOCODE Best Paper nomination, the FQRNT Postdoctoral Fellowship Award (2016), the FQRNT Doctoral Research Award (2012), and Best Paper Award at the 8th IEEE International NEWCAS Conference (2011). He is a member of IEEE, ACM, AAAS, IEEE Technical Committee on Cyber-Physical Systems (CPS), and the Editorial board member of the ACM Transactions on Probabilistic Machine Learning (TOPML) journal.